# Microscopic Polarization in Bilayer Graphene

Gregory M. Rutter[1*], Suyong Jung[1,2,*], Nikolai N. Klimov[1-3], David B. Newell[3], Nikolai B. Zhitenev[1§], and Joseph A. Stroscio[1§]

[1]Center for Nanoscale Science and Technology, NIST, Gaithersburg, MD 20899
[2]Maryland NanoCenter, University of Maryland, College Park, MD 20742
[3]Physical Measurement Laboratory, NIST, Gaithersburg, MD 20899

**Bilayer graphene has drawn significant attention due to the opening of a band gap in its low energy electronic spectrum, which offers a promising route to electronic applications. The gap can be either tunable through an external electric field or spontaneously formed through an interaction-induced symmetry breaking. Our scanning tunneling measurements reveal the microscopic nature of the bilayer gap to be very different from what is observed in previous macroscopic measurements or expected from current theoretical models. The potential difference between the layers, which is proportional to charge imbalance and determines the gap value, shows strong dependence on the disorder potential, varying spatially in both magnitude and sign on a microscopic level. Furthermore, the gap does not vanish at small charge densities. Additional interaction-induced effects are observed in a magnetic field with the opening of a subgap when the zero orbital Landau level is placed at the Fermi energy.**

Bilayer graphene consists of two graphene sheets overlaid in the Bernal stacking orientation where $A_2$ atoms of the top layer lie on top of the $B_1$ atoms of the bottom layer (see

---

[*] These authors contributed equally to this work.
[§] To whom correspondence should be addressed: nikolai.zhitenev@nist.gov, joseph.stroscio@nist.gov.

Fig. 1a), connected by the interlayer coupling $\gamma_1$, thus breaking the *A*/*B* sublattice symmetry in the individual graphene layers. This results in massive chiral fermions where the electronic energy dispersion is hyperbolic in momentum, in contrast to the linear dispersion that leads to massless carriers in single layer graphene[1,2]. In bilayer graphene the energy bands still meet at the charge neutrality point ($E_D$) in the absence of an electric field between the layers (neglecting interaction effects) (Fig. 1b). In an applied electric field a potential asymmetry is developed between the layers, resulting in the opening of an energy band gap between the low lying bands making bilayer graphene of intense interest in electronic applications (Fig. 1b)[2-10]. Bilayer graphene also differs from single layer graphene in its magnetic quantization in the quantum Hall regime. At $E_D$, the four-fold degenerate Landau level (LL) in single layer graphene becomes eight-fold degenerate in the bilayer due to the additional layer degeneracy[3,11,12]. When the gap is opened this manifold splits into two four-fold degenerate quartets polarized on each layer at low energies. Lifting of these degeneracies have been observed in recent measurements[13-16]. Theoretical studies[17,18] suggest the existence of interaction-driven band gaps, which are even possible in zero applied field with corresponding quantum Hall ferromagnetic states[17,19].

The energy band gap in bilayer graphene has been studied by optical measurements such as angle resolved photoemission spectroscopy[20] and infrared spectroscopy[4-6,21], which demonstrate that the gap is externally tunable and can reach values up to ≈ 250 meV. However, band gaps determined from conventional electronic transport measurements are smaller than theoretically expected or extracted from the optical measurements by an order of magnitude or more[7,8,13,22-25]. Recently, it has been suggested that disorder-induced localized states inside devices or along the edges can introduce additional conducting channels inside the gap, consequently reducing the effective gap seen in transport measurements[23-25]. Additionally, many-body interactions in bilayer graphene, which can be sensitive to local electrostatic variations such as disorder potential fluctuations, are expected to open a gap even in zero applied field[11,17]. Accordingly, the interplay between the interactions, external and disorder-induced local electric fields, and localized states in the gap is becoming the central issue in the physics of the bilayer graphene system. Direct atomic-scale probing with scanning tunneling microscopy (STM) and scanning tunneling spectroscopy (STS) has been proven as a powerful technique[26-28] for studying this physics, particularly in revealing the effects of disorder on the graphene electronic states[29-31].

In this article, we present the first STM/STS measurements of a gated bilayer graphene device in magnetic fields ranging from zero to the quantum Hall regime. We investigate the local density of states and the formation of an energy band gap affected by disorder while tuning the total charge density, as the Fermi energy ($E_F$) is varied with an electrostatic back gate with respect to $E_D$. Quite surprisingly, the determined local potential difference between the layers, which defines the gap, does not follow the previously reported electrostatic models[2,7] which predict that an external electric field is the main parameter in controlling the bilayer potential asymmetries. We observe the spatial variation of the potential difference to be highly correlated with the disorder potential. The potential difference between the layers reverses in sign between disorder potential minima and maxima locations, resulting in a pattern of oscillating charge imbalance between the graphene layers. We also observe a splitting of the zero orbital LL when it is placed at $E_F$ with an effective g-factor of ≈ 30, which we associate with correlated electron behavior in bilayer graphene. As a result, our experiment provides the first microscopic determination of a bilayer band gap exposing the major roles of the spatial varying disorder potential, the charge density variations that accompany LL filling, and many-body interactions.

The experimental setup is shown schematically in Fig. 1c. The experiments were performed on a graphene device fabricated on $SiO_2$/Si substrate by mechanical exfoliation and stencil mask evaporation (see methods below). The graphene sample contains both single and bilayer graphene, as seen in the STM topographic image in Fig. 2a. The topographic height fluctuations on the bilayer (Fig. 2b) are dominated by the underlying $SiO_2$ surface roughness, as they are in the single layer[29]. We have obtained the spatial profile of the bilayer disorder potential as shown in Fig. 2c (see methods below), over the 100 nm × 100 nm topographic region in Fig. 2b. The red and blue colored areas are the regions of low and high disorder potential that correspond to electron and hole puddles at near-zero carrier density, respectively. For brevity, we will refer to the disorder extrema as electron and hole puddles at arbitrary carrier densities for the rest of this manuscript. We note that the spatial size of the puddles in the bilayer is significantly smaller compared to the single layer with the same impurity densities[29], ≈ 10 nm in the bilayer compared to ≈ 30 nm in single layer. According to calculations[32], the smaller puddle size is caused by the increased screening properties of the bilayer.

We first discuss the measurements of the bilayer gap in zero magnetic field, followed by the measurements at high magnetic fields in both electron and hole puddles. Figure 3a shows a sequence of d$I$/d$V$ spectra at gate voltages ranging from 0 V to 60 V, obtained in the electron puddle denoted by P1 in Fig. 2c. In Fig. 3b, we use color-coded gate maps to plot the STS spectra obtained in finer gate potential steps, as previously applied to single layer graphene[29]. At $V_g$ = 0 V, we observe two main minima in the tunneling spectra (marked with red and orange triangles in Fig. 3a), one at $E_F$ and one at 80 mV. The gap at $E_F$ is characteristic to tunneling into low-dimensional systems, and we associate the minimum centered at 80 mV with the band gap of bilayer graphene. Because of the multiple peaks seen in the sequence of spectra in Fig. 3a, which we relate to scattering resonances in the disorder potential[29], the reliable assignment of the bilayer band gap is only possible after careful study of the zero field gate map in Fig. 3b *and* the magnetic field dependent measurements, such as the ones displayed in Fig. 3e–g. The minima associated with the bilayer gaps are observed to increase in energy width as a function of magnetic field (Fig. 3e–g) with the development of Landau levels, as expected (see discussion below). Once the gap is identified, it can be tracked as a function of gate voltage as shown by the orange triangles in Fig. 3a and green circles in Fig. 3b. Additionally, we determine the edges of the gap as the closest peaks on either side of the gap minima (indicated by the red and blue dots in Fig. 3b).

The charge neutrality point can be extracted from the center of the gap (green circles in Fig. 3b), which varies linearly with density as $E_D = \hbar^2 \pi n / 2m^*$, in contrast to the square-root dependence in single layer[29]. Here, the two-dimensional charge-carrier density $n$ is defined by the applied gate potential, $n = \alpha (V_g - V_o)$, where $\alpha$ is determined by the gate capacitance (300 nm of $SiO_2$) as $7.19 \times 10^{10}$ cm$^{-2}$ V$^{-1}$ and $V_o$ is the shift of $E_D$ due to intrinsic doping[26,29,33]. A linear fit (yellow line in Fig. 3b) to $E_D$ with density yields an effective mass $m^* = (0.033 \pm 0.002)\, m_e$[34], where $m_e$ is the mass of electron, in agreement with bilayer graphene properties[35]. Using the interlayer coupling constant, $\gamma_1 = 2m^* v_F^2 = 0.377$ eV [35], we can extract the Fermi velocity as $v_F = (1.010 \pm 0.003) \times 10^6$ m s$^{-1}$. The charge neutrality point is close to $E_F$ in this puddle at a gate voltage of $V_g \approx 30$ V. Spatially, $V_o$ varies from $V_g \approx 30$ V (electron puddles) to $V_g \approx 35$ V (hole puddles).

We now discuss the bilayer spectrum in the quantum Hall regime at high magnetic fields. In the presence of a perpendicular magnetic field $B$, the massive chiral fermions in gapless bilayer graphene are quantized with energies $E_N = \pm\hbar\omega_c\sqrt{N(N-1)}$, $N = 0,1,2,\cdots$, where $\hbar$ is Planck's constant, $\omega_c = eB/m^*$ is the cyclotron frequency, and $e$ is the electron charge[12,36]. For each orbital quantum number, $N$, the Landau levels ($LL_N$) are four-fold degenerate due to the degeneracy of valleys $\boldsymbol{K}$ and $\boldsymbol{\tilde{K}}$ with respective quantum numbers $\xi = +1$ and $\xi = -1$, and spin degeneracy, $s = \uparrow, \downarrow$. In the absence of an applied electric field and interactions, the $N = 0$ and $N = 1$ LLs are degenerate and an eight-fold degeneracy occurs at $E_{N=0,1} = 0$ meV. With an applied electric field this degeneracy is partially lifted and a band gap is opened in the low energy bands. Accordingly, the Landau level energies are modified as[3,11,12],

$$E_N = \pm\hbar\omega_c\sqrt{N(N-1)+\left(\Delta U/2\hbar\omega_c\right)^2} - \frac{1}{4}\xi z \Delta U, \quad N = 2,3,4\cdots,$$
$$E_{N=0} = \xi\Delta U/2, \quad E_{N=1} = \xi(\Delta U/2)(1-z)$$
(1)

with the potential energy difference (or asymmetry) $\Delta U = U_2 - U_1$, where $U_2$ and $U_1$ are the on-site energies of the top and bottom graphene layers, respectively (Fig. 1c), and $z$ is a term relating to the $\boldsymbol{B}_2/\boldsymbol{A}_1$ dimer sites given by $z = 2\hbar\omega_c/\gamma_1 \ll 1$. For $\Delta U < \gamma_1$, the absolute value of the potential energy asymmetry is approximately equal to band gap, $|\Delta U| \approx E_g$ (see Fig. 1b). As $z$ is small for typical magnetic fields (B ≤ 8 T), the $N = 0$ and $N = 1$ LLs are nearly degenerate. We label these states as $LL_{(N,\xi=\pm)}$ with orbital and valley indices $N$ and $\xi = \pm$, without the spin index label. Landau levels that are degenerate at different orbital indices ($N$ and $N'$) are separated by a semicolon in the notation as $LL_{(N,\pm);(N',\pm)}$.

Significantly, the spinor states related to the $N = 0$ and $N = 1$ states in the $\boldsymbol{\tilde{K}}$ ($\xi = -1$) valley ($LL_{(0,-);(1,-)}$) are localized predominantly on the $\boldsymbol{A}_1$ sites of the bottom layer and the $N = 0$ and $N = 1$ states in the $\boldsymbol{K}$ ($\xi = +1$) valley ($LL_{(0,+);(1,+)}$) are located on the $\boldsymbol{B}_2$ sites of the top layer[11,12]. This layer polarization is of particular importance in the scanning tunneling spectroscopy measurements, as tunneling to the top surface layer dominates in d$I$/d$V$ spectra. Analysis of the state, $LL_{(0,+);(1,+)}$ that resides predominantly in the top layer and belongs either to the valence band ($\Delta U < 0$) or to the conduction band ($\Delta U > 0$) (Fig. 1d) enables us to determine both the sign and the value of $\Delta U$. Each quartet of the $LL_{(0,\pm);(1,\pm)}$ manifold remains degenerate at each layer, but electron-electron interactions can further lift this degeneracy and enhance the splitting of this level as pointed out in theoretical analyses[11,17,19].

A rich set of spectral features is observed in the STS spectra in bilayer graphene in the quantum Hall regime. Figure 3c shows the gate map at the location of the P1 electron puddle (Fig. 2c) at 8 T. Resonance peaks from impurity scattering become suppressed as graphene charged carriers are condensed into LLs. The bright lines in Fig. 3c represent well-defined LLs, observed up to $LL_{N=6}$ at both electron and hole doping. A staircase-like pattern is observed in the LL transitions as a function of gate voltage, which results from the LLs pinning at $E_F$[29], and is characteristic of a 2DEG in high magnetic fields[37]. A large gap in the d$I$/d$V$ spectrum is observed in the gate map (Fig. 3c) as a function of gate voltage with two prominent LLs on either side of the gap. For comparison, the band gap edges observed at zero magnetic field are overlaid on the map (red and blue dots) in Fig. 3c. To identify the orbital index $N$ of each LL, we examine the magnetic field dependence of the d$I$/d$V$ spectra and LL peak positions as shown in Fig. 3e. In the electron puddles, a prominent LL is observed to grow out of the lower energy side of the gap with minimal dispersion in magnetic field. This LL belongs to the valence band, and we identify this level as $LL_{(0,+);(1,+)}$ which resides in the top graphene layer at high magnetic field. The corresponding level $LL_{(0,-);(1,-)}$ residing on the bottom layer, is not observed as the tunneling probability from the probe tip to this layer is exponentially decreased ($\approx e^{-\kappa \Delta Z} \approx e^{-7}$ with the decay constant $\kappa = 20$ nm$^{-1}$ and interlayer separation $\Delta Z = 0.35$ nm). The same level assignment is valid for the whole gate voltage range. $LL_{(0,+);(1,+)}$ emerging from the valence band can also be seen by comparing the gap edge positions at zero field (red and blue dots in Fig. 3c) with the LLs at high field in the gate map at 8 T. With the $LL_{(0,+);(1,+)}$ quartet assigned, the other LLs at higher orbital indices can be indexed accordingly as marked in Fig. 3e.

The bilayer band gap varies dramatically on a microscopic level, which was not anticipated before this study. Figure 3d shows the spectral gate map measured at the position of a hole puddle ('P4' in Fig. 2c), which can be contrasted with the measurements in an electron puddle ('P1' in Fig. 3c). The spectra in the electron and hole puddles are reproducibly distinct, as can be seen from the comparison of another two electron and hole puddles in Fig. 4b,c. Similar to single layer graphene[29], the gate maps in the hole puddles (Fig. 3d and 4b) show LL transitions that display convex curvature when the LLs are pinned at $E_F$, compared to the concave transitions observed in the electron puddles (Fig. 3c and 4c). In comparing the spectral peaks in the electron vs. hole puddles we observe a striking difference in the layer-polarized LLs. As shown in Fig. 3f, the strong non-dispersive $LL_{(0,+);(1,+)}$ grows out of the higher energy side of the bilayer gap at the conduction band edge (blue dots in Fig. 3d), as opposed to the $LL_{(0,+);(1,+)}$ growing out of the valence band for electron puddles (Fig. 3e). This indicates a *reversal* in the sign of the electric field between the graphene layers, which results in a sign change of the energy asymmetry, $\Delta U$ (see Fig. 1c).

From the LL spectral peak positions, we can quantitatively determine the values of the bilayer energy gap in the different puddles from Eq. 1 using the energy asymmetry $\Delta U$ and the Fermi-velocity $v_F$ as fitting parameters. The dark-brown tick marks in Fig. 3e (Fig. 3f) indicate the fitted LL positions from Eq. 1 using an energy asymmetry $\Delta U = -34.8$ meV ($\Delta U = 31.9$ meV), and Fermi velocity $v_F = 1.00 \times 10^6$ m s$^{-1}$ ($v_F = 1.01 \times 10^6$ m s$^{-1}$) for the electron (hole) puddle P1 (P4). The fan diagram (solid lines in Fig. 3g) generated with the fit parameters for the P1 puddle shows how individual LLs are expected to evolve as a function of magnetic field, along with experimental LL energies for the field range from 2 T to 8 T and the band gap edges determined at 0 T. Fairly good agreement is observed over the entire field range with the gap and velocity fixed to the values obtained by fitting the 8 T spectra. Better agreement between the model and measured LL energies can be obtained by fitting the energy asymmetry and velocity for each magnetic field.

The extracted $\Delta U$ values at 6 T and 8 T as a function of gate voltage are shown in Fig. 4a together with $\Delta U$ measured in zero field (red squares) for the electron puddle P1 and hole puddle P4. We note that qualitatively $\Delta U$ follows the separation between $LL_{(0,+);(1,+)}$ and $LL_{(2,+);(2,-)}$, which can be seen directly in the gate map in Fig. 3c,d (see also supplemental material). Interestingly, the $\Delta U$ dependence on density shows an almost mirror symmetry about zero energy for electron ($\Delta U < 0$) and hole ($\Delta U > 0$) puddles. The magnitude of $\Delta U$ is comparable at opposite ends of the doping range. The energy difference measured in the electron puddle at $V_g$ = 0 V is ≈ -35 meV. For the hole puddle, the energy difference is ≈ +35 meV at $V_g$ = 60 V. Furthermore, it is interesting to note that the energy difference at low gate voltage (0 V < $V_g$ < 20 V) follows the energy difference measured in zero applied field for the electron puddle. For the hole puddle, however, the energy difference in higher magnetic fields match with those in zero field at high gate voltage (40 V < $V_g$ < 60 V). Additionally, there exist a series of dips (peaks) in the electron (hole) energy asymmetries at higher (lower) gate voltage, each corresponding to the transitions of the various LLs through $E_F$. The peak at $V_g$ = 27 V in hole puddle data corresponds to the filling of $LL_{(2,+);(2,-)}$. In comparison, the dip at $V_g$ = 40 V in electron puddle corresponds to the filling of $LL_{(2,+);(2,-)}$ and the dip at $V_g$ = 50 V corresponds to the filling of $LL_{(3,+);(3,-)}$ at $E_F$.

Coincident with the variation in the gap size is a few-percent variation in the value of the Fermi velocity (see supplemental material). We attribute these variations to velocity renormalization effects due to electron-electron interactions[38].

We compare the observed density dependence of $\Delta U$ with the zero-magnetic field calculations (blue line in Fig. 4a) using a tight-binding model with a self-consistent Hartree approximation[2]. The model predicts a vanishing $\Delta U$ and sign reversal at $E_D$. The models[11,12] describing the bilayer band structure in high magnetic field are not self-consistent and implicitly assume an energy asymmetry dependence on the gate electric field illustrated by the dashed blue line in Fig. 4a. Here, the energy gap opens and the layer polarization develops when $E_D$ crosses $E_F$, while LLs with higher orbital indices, $N \geq 2$, are not layer polarized and therefore do not contribute to changes in relative charge imbalance or the asymmetry size. In contrast, our observations show strong peaks and dips in the energy difference when higher orbital LLs are filled or emptied. It is clear that the experimental observations presented in this manuscript are very different from what either model predicts.

A gap (or subgap) of another type is seen as a splitting of the $LL_{(0,+);(1,+)}$ when it crosses $E_F$. The splittings have been observed in all of six different electron puddle locations examined, marked in Fig. 2c. The splitting of $LL_{(0,+);(1,+)}$ at $E_F$ is also seen as resonances that appear as nearly vertical lines in the gate maps (see Fig. 4c and Fig. 5a). The presence of vertical resonances in gate maps was discussed in a recent study on single layer graphene[29], where we showed that the physical phenomena at $E_F$ can also contribute to the d$I$/d$V$ spectra at higher tunneling energies. The resonances occur when the split $LL_{(0,+);(1,+)}$ levels are pulled through the Fermi level at high tip-sample potentials giving a step increase in tunneling current and a resonance peak in the d$I$/d$V$ measurements[39]. The leftmost resonance corresponds to the transition between filling factors of -4 to -2 at $E_F$, and the rightmost resonance corresponds to the transition between filling factors of -2 to 0.

The splitting of $LL_{(0,+);(1,+)}$ at $E_F$ is a sign of correlated electron behavior[28]. We examine this splitting in more detail in Fig. 5b. The inset in Fig. 5b shows an individual d$I$/d$V$ spectrum for the electron puddle P2 in the middle of the subgap, at $V_g = 28.6$ V. The four-fold degeneracy of $LL_{(0,+);(1,+)}$ is partially lifted and it splits into two peaks separated by 15.4 meV at 8 T. The gap is nearly constant and collapses suddenly when $LL_{(0,+);(1,+)}$ is moved away from $E_F$. The peak separation scales linearly with magnetic field, with slight variation on different puddles as shown in Fig. 5b. Fitting the splitting energies to Zeeman-like dependence, $E = g\mu_B B$, yields an energy scale for the splitting that is extremely large, $\approx 1.97$ meV T$^{-1}$, with effective $g \approx 34$ for the puddle P2 and $\approx 1.70$ meV T$^{-1}$ with $g \approx 29$ for the puddle P1. Interestingly, these subgaps are not resolved in the hole puddles (see Fig. 4b), implying that the splitting may be much smaller there.

In the following discussion, we would like to emphasize that the experimental results cannot be explained by models considering a spatially homogeneous layer polarization that goes to zero and reverses in sign when $E_F$ passes through $E_D$ [2,7] (blue line in Fig. 4a). As discussed previously, the measured bilayer band gap remains open with values on the order of 25 meV even when $E_D$ coincides with $E_F$ (variations depend on spatial position in the disorder potential). Our measurements demonstrate that both the total charge density and the charge imbalance between the layers spatially fluctuate reflecting the disorder potential variation. Moreover, the direction of electric field between the layers, the deciding factor to determine the sign of the energy asymmetry, is observed to also fluctuate with spatial locations even in the presence of large applied gate fields.

Quite surprisingly, therefore, the direction of the local electric field between the layers determined by the sign of the charge difference remains unchanged over the whole experimentally explored density range (0 V < $V_g$ < 60 V), as schematically illustrated in Fig. 4d, even though the total electric field (applied gate field plus the disorder-induced field) and the carriers change sign when $E_D$ crosses $E_F$. The total electric field defines the total charge density, which is proportional to the applied gate voltage, as seen, for example, in the linear $E_D$ variation with gate potential (Fig. 3b). The charge density range controlled by the external gate ($\Delta V_g$ = 60 V corresponds to $\Delta n = \pm 2.2 \times 10^{12}$ cm$^{-2}$)[29], is significantly larger than the density fluctuations between minima and maxima potential extrema ($\Delta V_o$ = 5 V corresponds to $\Delta n = 3.6 \times 10^{11}$ cm$^{-2}$)[29], implying that the applied gate electric field is significantly larger than the disorder-induced field. However, the sign of the local polarization remains fixed in the respective puddles (Fig. 4d).

As graphically illustrated in Fig. 4d, at overall hole doping in the bilayer (leftmost panel), the electron puddles have a large energy asymmetry (red arrow) while the asymmetry in the hole puddles becomes small (blue arrow). Here, the sign of the local polarization field coincides with the external field in electron puddles and is opposite in the hole puddles. The opposite trend occurs at electron doping in the bilayer consistent with the reversal of the external electric field at $V_g \gg V_D$ (rightmost panel in Fig. 4d). Importantly, the direction of the local polarization field does not follow the direction of the external electric field and must be determined by other factors such as the gradients of the field that change sign in different puddles. The schematic also offers a possible simple clue. Over the whole density range, the amplitude of density fluctuations in the bottom layer is larger than the one in the top layer consistent with the screening of the substrate-induced potential disorder. The resulting density schematics shown as the line profiles in Fig. 4d naturally result in the potential asymmetries and local fields reversing in sign from electron to hole puddles.

Even though we observe a non-zero energy gap when $E_D$ crosses $E_F$, it is not clear whether the observed asymmetries are related to the broken symmetry states predicted in recent models[11,17,19,40,41], or are the result of the broken symmetry related to the substrate interactions. In contrast, the existence of the subgaps in electron puddles when $LL_{(0,+);(1,+)}$ crosses $E_F$ is likely related to the recent theoretical predictions of spontaneously broken symmetry states, as correlated electron behavior is expected and most easily observed when the LLs are close to $E_F$ [28]. At present, we are not able to identify the exact quantum numbers of the split $LL_{(0,+);(1,+)}$ levels. However, recent transport measurements point to a pseudospin polarized ground state[13,19,40]. The spontaneously broken symmetry states lead to the opening of a gap due to many-body interactions in zero electric field[13,17,19,40]. In this model, a ferromagnetic ground state of pseudospin polarized states is favored at small electric fields, which is followed by antiferrromagnetic or ferrimagnetic ordering. These effects lead to non-monotonic dependence of the band gap on applied electric field with magnitudes comparable to those observed in this study[17]. However, in realistic devices with disorder, the spontaneous polarization must nucleate with a sign determined by the disorder potential variation.

**Methods**

The experiments were performed with an ultra-high vacuum (UHV) STM facility at NIST in magnetic fields from 0 T to 8 T at a temperature of 4.3 K. The graphene device was fabricated in a similar way to that reported in Novoselov et al.[42]. Graphene flakes were mechanically exfoliated from natural graphite and transferred on thermally grown 300 nm thick $SiO_2$ on Si.

The highly doped Si substrate was used as a back gate to control the charge density of the graphene device. Multiple steps of gold evaporation (50 nm for single deposition) through a SiN stencil mask were implemented to preserve a clean surface of graphene. Raman spectroscopy measurements were performed to determine the single and bilayer graphene regions[29]. The graphene was located using a 2-dimensional piezoelectric actuator to position the probe tip on the graphene device using optical viewing. An Ir probe tip was prepared by ex-situ electrochemical etching, and cleaned and characterized by in-situ field ion microscopy before the measurements. STS measurements were performed using a lock-in detection method with a modulation frequency of ≈ 500 Hz and root-mean-square modulation voltages between 1 mV and 8 mV depending on the spectral range of interest. The disorder potential map in Fig. 2c was obtained using closed loop d$I$/d$V$ measurements as described previously[29,31].

**Acknowledgments** We would like to acknowledge M. Stiles, S. Adam, H. Min, and A. H. MacDonald for fruitful discussions and thank I. Calizo and A. Hight-Walker for Raman spectroscopy characterization of the graphene system.

**Author contributions** The graphene sample was fabricated by S.J. and N.N.K. STM/STS measurements were performed by G.M.R., S. J., N.N.K. and J.A.S. The data analysis and preparation of the manuscript were performed by G.M.R., S. J., J.A.S., D.B.N and N.B.Z.

**Additional information** The authors declare no competing financial interests. Supplementary information accompanies this paper on www.nature.com/naturephysics. Reprints and permissions information is available online at http://npg.nature.com/reprintsandpermissions. Correspondence and requests for materials should be addressed to N.B.Z and J.A.S.

**Figure Captions**

**Figure 1: Schematics of the bilayer graphene measurements and energy band diagram in the quantum Hall regime. a**, Schematic of Bernal-stacked bilayer graphene consisting of a top layer ($A_2/B_2$) and bottom layer ($A_1/B_1$), with atom $A_2$ directly over $B_1$. **b**, Energy band diagram of bilayer graphene with (solid lines) and without (dashed lines) a band

gap. The electronic levels form a Mexican-hat like energy bands with a potential energy asymmetry $\Delta U$ and a band gap of $E_g$. **c**, Schematic of a gated bilayer graphene device for STM/STS measurement with circuitry showing application of gate voltage $V_g$ and sample bias $V_b$. The bilayer graphene is placed on a 300 nm $SiO_2$ substrate separating from a back gate electrode (Si). The disorder potential induced from the substrate is illustrated in color overlaid on the $SiO_2$ surface. $\Delta U$ equals to the difference between onsite energies for the top (2) and the bottom (1) layers. Magnetic field $B$ is perpendicular to the sample plane. **d**, The formation of bilayer graphene Landau levels in the quantum Hall regime with and without a band gap. Landau levels are indexed with the orbital and valley index, $(N, \xi)$, and each is two-fold degenerate in spin. The eight-fold degenerate $N = 0, 1$ levels become layer polarized quartets when the graphene layers are subjected to a potential energy asymmetry $\Delta U$. $LL_{(0,+);(1,+)}$ projected on the top layer ($\xi = +1$) depends on the sign of $\Delta U$.

**Figure 2: STM topography images and disorder potential in bilayer graphene. a**, STM topographic image, 200 nm × 200 nm, of a region containing the boundary between single and bilayer graphene. Tunneling parameters: sample bias -300 mV, tunneling current 100 pA. Lower right inset: atomic resolution image of the honeycomb lattice structure of single layer graphene. Lower left inset: atomic resolution image of the bilayer showing the three-fold symmetry of Bernal-stacked bilayer graphene. **b**, STM topographic image, 100 nm × 100 nm, of bilayer graphene with peak-to-peak height corrugation of 1 nm. Tunneling parameters: sample bias -200 mV, tunneling current 300 pA. **c**, Fixed-bias closed-loop d$I$/d$V$ map ($V_b$ = -200 mV, $V_g$ = 60 V) over the same area as in **b** revealing the spatial distribution of the disorder potential in bilayer graphene. Different measurement points, six for electron and two for hole puddles are indicated.

**Figure 3: Magnetic quantization in bilayer graphene as a function of electric and magnetic fields. a**, d$I$/d$V$ spectra in zero magnetic field as a function of back gate voltage in steps of $\Delta V_g = 5$ V in the electron puddle P1. The curves are offset for clarity. The red tick mark at $V_g = 0$ V indicates the conductance minimum at $E_F$ while the orange tick marks indicate the minima of the bilayer band gap as a function of gate voltage. **b**, 2-dimensional 'gate map' from d$I$/d$V$ spectra with fine gate voltage increments ($\Delta V_g = 0.2$ V). The green circles show the position of the charge neutrality point, $E_D$, in the center of the gap and the yellow line is a linear fit of $E_D$ vs. gate voltage. The red and blue connected dots denote the edges of band gap. **c**, d$I$/d$V$ gate map measured at 8 T at the same location as **a** and **b** (P1). The red and blue connected dots are the same as in **b**. **d**, d$I$/d$V$ gate map measured at 8 T at the position of hole puddle P4. The red and blue connected dots denote the gap edges of the hole puddle determined from zero field measurements. **e-f**, Individual d$I$/d$V$ spectra for the electron puddle P1 and the hole puddle P4 as a function of applied magnetic field from 0 T to 8 T at the fixed gate voltage of $V_g = 33$ V, respectively. The curves are offset for clarity. Dark brown tick marks show calculated LLs positions for each puddle from Eq. 1 in the main text. STS parameters (**a-f**): sample bias -200 mV, tunneling current 200 pA and root-mean-square modulation voltage 4 mV. **g**, Landau level peak positions (red squares) as a function of magnetic field for the electron puddle P1 at $V_g = 33$ V (error bars less than symbol size). The experimental points at 0 T and 2 T are extracted from the band gap edges.

**Figure 4: Bilayer graphene potential energy asymmetries at varying gate voltage in electron vs. hole puddles. a**, Energy asymmetries determined from fitting the LLs using Eq. 1 as a function of gate voltage and magnetic field obtained at the spatial location of the electron puddle P1 and hole puddle P4. Error bars are one standard deviation[34]. The solid blue line indicates the calculated potential asymmetry using Eq. 5 in reference 9 with $\Delta_0 = -76$ meV, $v_F = 1.00 \times 10^6$ m s$^{-1}$, and $\varepsilon_r = 1$. The dashed blue line is the asymmetry dependence implied in the LLs scheme illustrated in Fig. 1d. **b–c,** Comparison of d$I$/d$V$ gate maps at 8 T measured in the hole puddle P3 (**b**) and electron puddle P8 (**c**). STS parameters: sample bias -200 mV, tunneling current 200 pA and root-mean-square modulation voltage 4 mV. **d,** Schematics of the spatial inhomogeneity of the layer densities in bilayer graphene at different gate potentials according to the observations made in **a**. The plots on the bottom illustrate the density variation in both top (dashed line) and bottom (solid line) layers along the direction marked by the yellow arrows above. The amplitude of density fluctuations is smaller in the top layer because of the screening from the bottom layer. The direction of electric field between the layers and the sign of energy asymmetry remain the same over the whole explored density range.

**Figure 5: Symmetry breaking in the $LL_{(0,+);(1,+)}$ quartet. a**, d$I$/d$V$ gate map at 8 T in electron puddle P2. $LL_{(0,+);(1,+)}$ is observed to split into two peaks opening a subgap when intersecting $E_F$ inside the dashed yellow rectangle. STS parameters: sample bias -200 mV, tunneling current 200 pA and root-mean-square modulation voltage 4 mV. **b**, The subgap energy vs. magnetic field for electron puddles P1 and P2. A linear fit of the gap vs. magnetic field yields the slopes of $(1.70 \pm 0.21)$ meV T$^{-1}$ and $(1.97 \pm 0.03)$ meV T$^{-1}$ for the positions P1 and P2, respectively. (Inset) d$I$/d$V$ spectra in the middle of the subgap of the electron puddle P2 at $V_g = 28.6$ V as marked in the gate map **a**, showing the subgap size of 15.4 meV at 8 T.

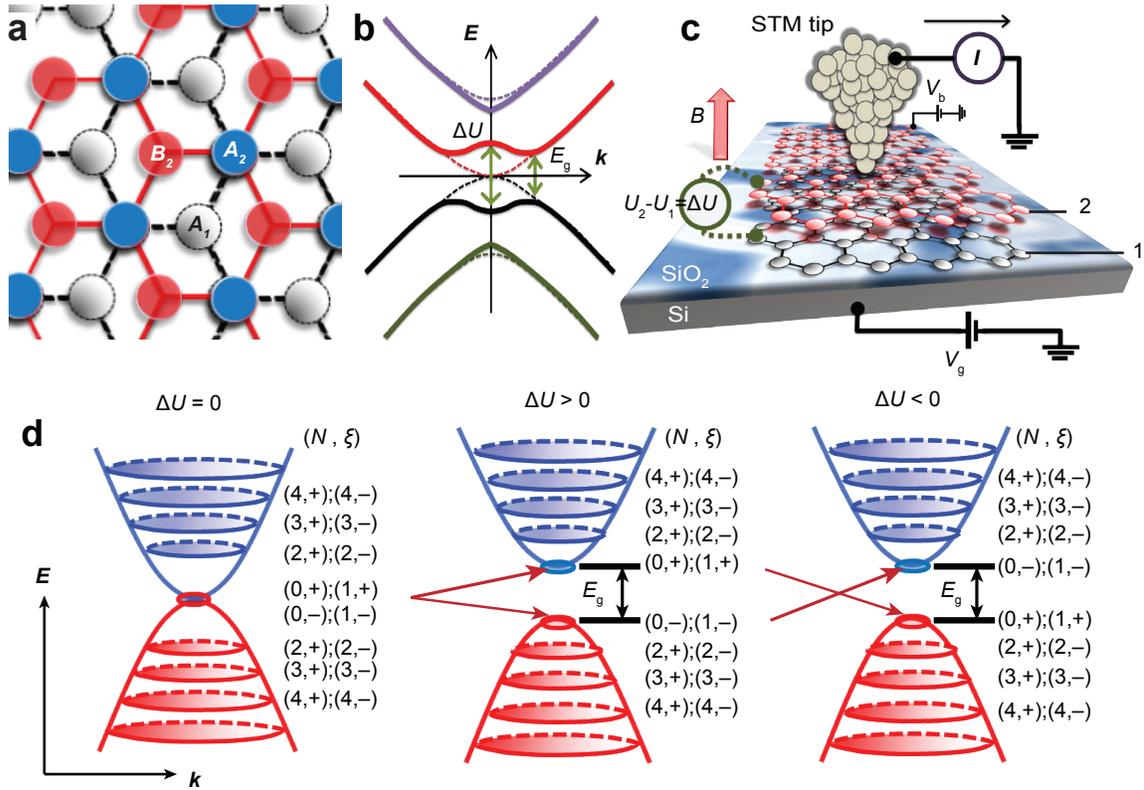

**Figure 1: Schematics of the bilayer graphene measurements and energy band diagram in the quantum Hall regime. a**, Schematic of Bernal-stacked bilayer graphene consisting of a top layer ($A_2/B_2$) and bottom layer ($A_1/B_1$), with atom $A_2$ directly over $B_1$. **b**, Energy band diagram of bilayer graphene with (solid lines) and without (dashed lines) a band gap. The electronic levels form Mexican-hat like energy bands with a potential energy asymmetry $\Delta U$ and a band gap of $E_g$. **c**, Schematic of a gated bilayer graphene device for STM/STS measurement with circuitry showing application of gate voltage $V_g$ and sample bias $V_b$. The bilayer graphene is placed on a 300 nm $SiO_2$ substrate separating from a back gate electrode (Si). The disorder potential induced from the substrate is illustrated in color overlaid on the $SiO_2$ surface. $\Delta U$ equals to the difference between onsite energies for the top (2) and the bottom (1) layers. Magnetic field $B$ is perpendicular to the sample plane. **d**, The formation of bilayer graphene Landau levels in the quantum Hall regime with and without a band gap. Landau levels are indexed with the orbital and valley index, $(N, \xi)$, and each is two-fold degenerate in spin. The eight-fold degenerate $N = 0, 1$ levels become layer polarized quartets when the graphene layers are subjected to a potential energy asymmetry $\Delta U$. $LL_{(0,+);(1,+)}$ projected on the top layer ($\xi = +1$) depends on the sign of $\Delta U$.

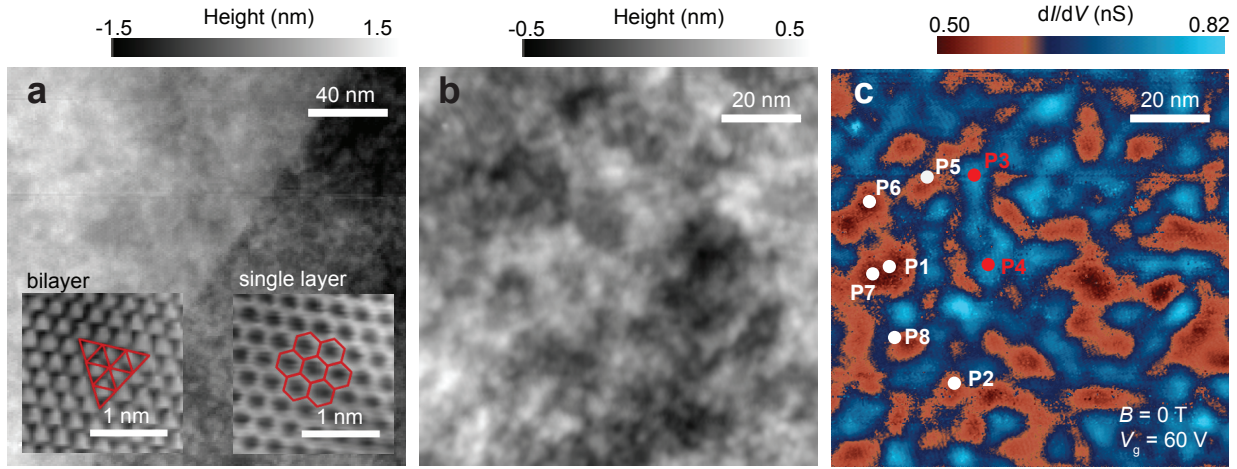

**Figure 2: STM topography images and disorder potential in bilayer graphene. a**, STM topographic image, 200 nm × 200 nm, of a region containing the boundary between single and bilayer graphene. Tunneling parameters: sample bias -300 mV, tunneling current 100 pA. Lower right inset: atomic resolution image of the honeycomb lattice structure of single layer graphene. Lower left inset: atomic resolution image of the bilayer showing the three-fold symmetry of Bernal-stacked bilayer graphene. **b**, STM topographic image, 100 nm × 100 nm, of bilayer graphene with peak-to-peak height corrugation of 1 nm. Tunneling parameters: sample bias -200 mV, tunneling current 300 pA. **c**, Fixed-bias closed-loop $dI/dV$ map ($V_b$ = -200 mV, $V_g$ = 60 V) over the same area as in **b** revealing the spatial distribution of the disorder potential in bilayer graphene. Different measurement points, six for electron and two for hole puddles are indicated.

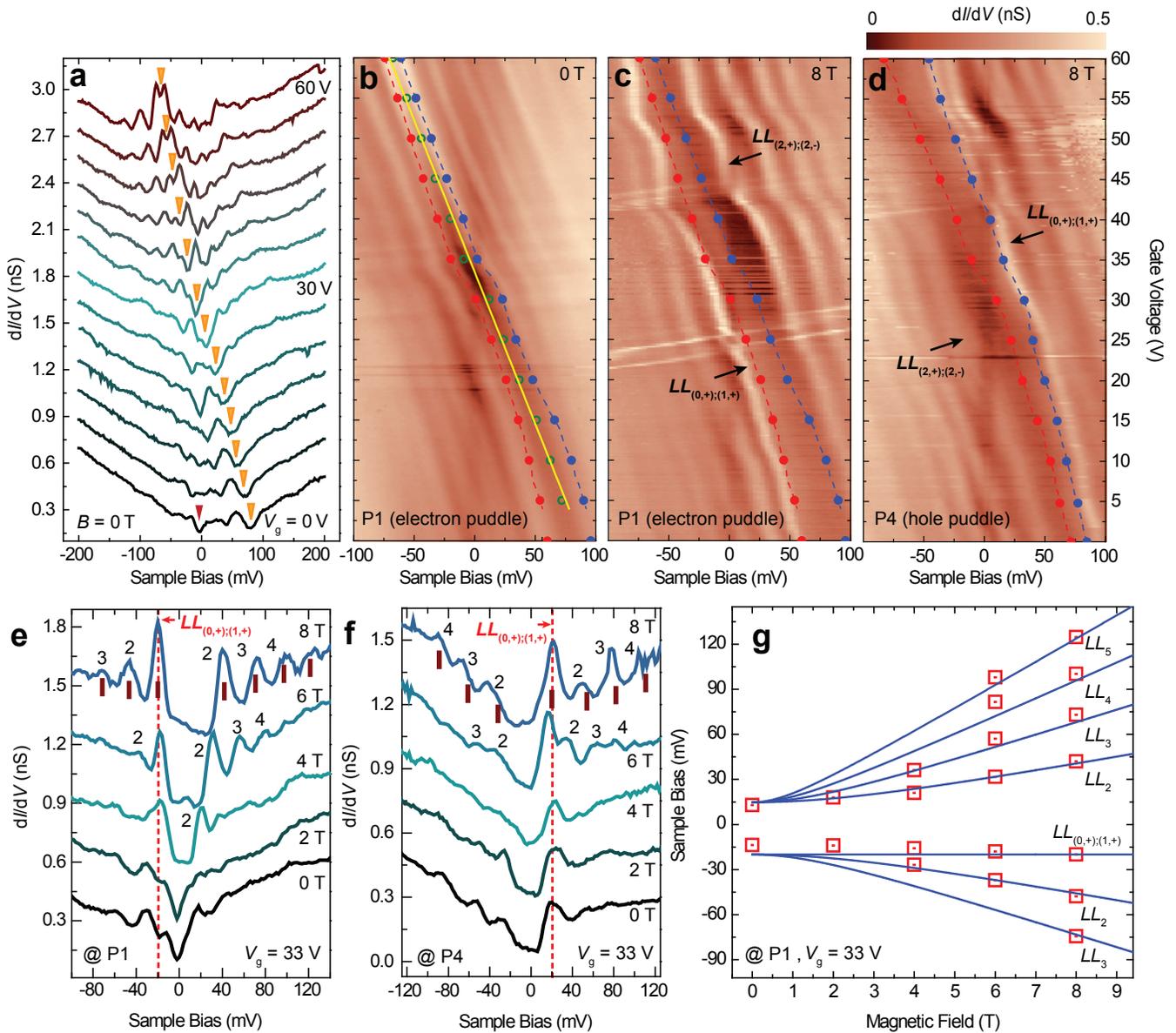

**Figure 3: Magnetic quantization in bilayer graphene as a function of electric and magnetic fields. a**, d$I$/d$V$ spectra in zero magnetic field as a function of back gate voltage in steps of $\Delta V_g = 5$ V in the electron puddle P1. The curves are offset for clarity. The red tick mark at $V_g = 0$ V indicates the conductance minimum at $E_F$ while the orange tick marks indicate the minima of the bilayer band gap as a function of gate voltage. **b**, 2-dimensional 'gate map' from d$I$/d$V$ spectra with fine gate voltage increments ($\Delta V_g = 0.2$ V). The green circles show the position of the charge neutrality point, $E_D$, in the center of the gap and the yellow line is a linear fit of $E_D$ vs. gate voltage. The red and blue connected dots denote the edges of band gap. **c**, d$I$/d$V$ gate map measured at 8 T at the same location as **a** and **b** (P1). The red and blue connected dots are the same as in **b**. **d**, d$I$/d$V$ gate map measured at 8 T at the position of hole puddle P4. The red and blue connected dots denote the gap edges of the hole puddle determined from zero field measurements. **e-f**, Individual d$I$/d$V$ spectra for the electron puddle P1 and the hole puddle P4 as a function of applied magnetic field from 0 T to 8 T at the fixed gate voltage of $V_g = 33$ V, respectively. The curves are offset for clarity. Dark brown tick marks show calculated LLs positions for each puddle from Eq. 1 in the main text. STS parameters (**a-f**): sample bias -200 mV, tunneling current 200 pA and root-mean-square modulation voltage 4 mV. **g**, Landau level peak positions (red squares) as a function of magnetic field for the electron puddle P1 at $V_g = 33$ V (error bars less than symbol size). The experimental points at 0 T and 2 T are extracted from the band gap edges.

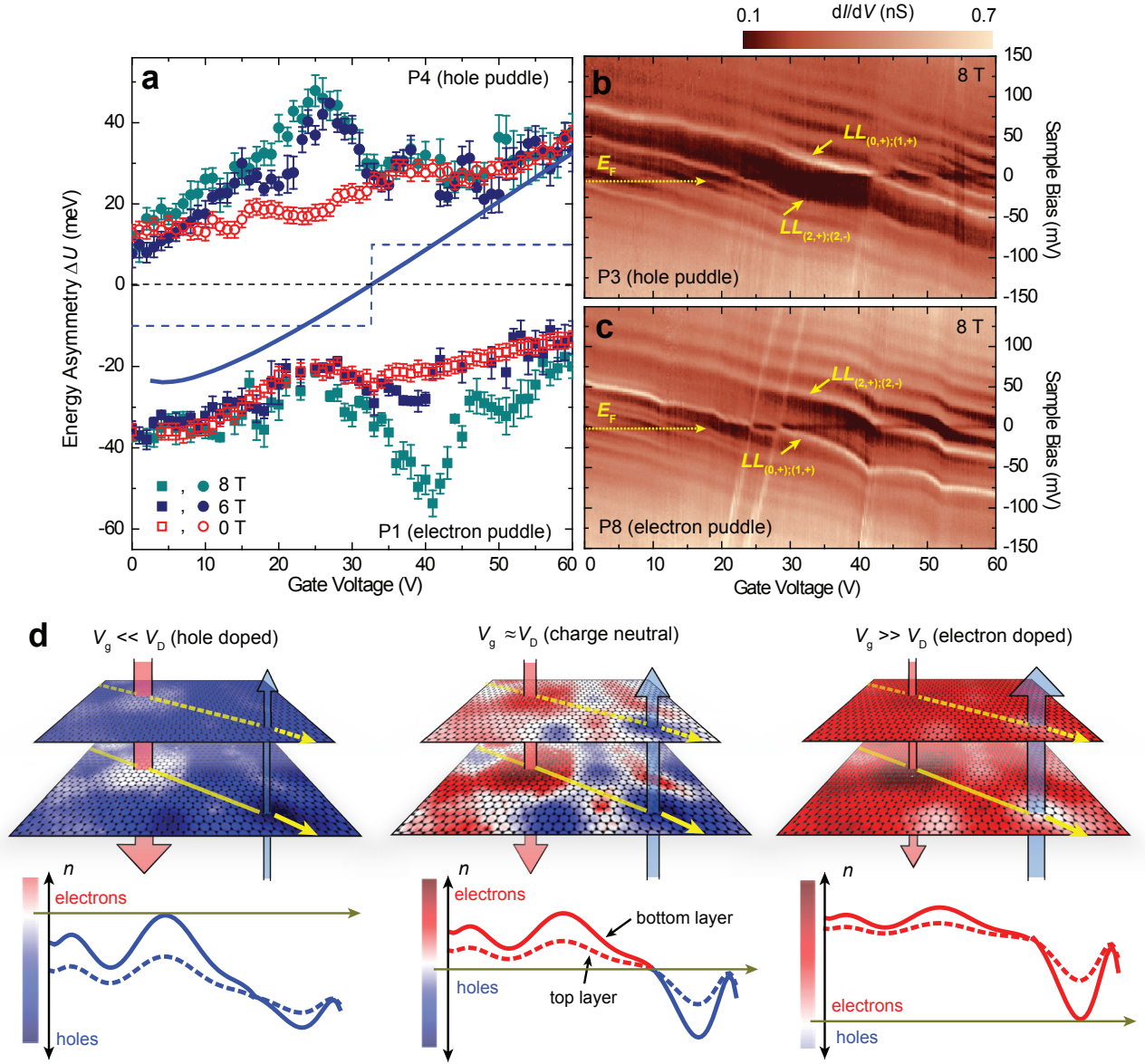

**Figure 4: Bilayer graphene potential energy asymmetries at varying gate voltage in electron vs. hole puddles. a**, Energy asymmetries determined from fitting the LLs using Eq. 1 as a function of gate voltage and magnetic field obtained at the spatial location of the electron puddle P1 and hole puddle P4. Error bars are one standard deviation[33]. The solid blue line indicates the calculated potential asymmetry using Eq. 5 in reference 2 with $\Delta_0$ = -76 meV, $v_F$ = 1.00 × 10$^6$ m s$^{-1}$, and $\varepsilon_r$ = 1. The dashed blue line is the asymmetry dependence implied in the LLs scheme illustrated in Fig. 1d. **b-c**, Comparison of d$I$/d$V$ gate maps at 8 T measured in the hole puddle P3 (**b**) and electron puddle P8 (**c**). STS parameters: sample bias -200 mV, tunneling current 200 pA and root-mean-square modulation voltage 4 mV. **d**, Schematics of the spatial inhomogeneity of the layer densities in bilayer graphene at different gate potentials according to the observations made in **a**. The plots on the bottom illustrate the density variation in both top (dashed line) and bottom (solid line) layers along the direction marked by the yellow arrows above. The amplitude of density fluctuations is smaller in the top layer because of the screening from the bottom layer. The direction of electric field between the layers and the sign of energy asymmetry remain the same over the whole explored density range

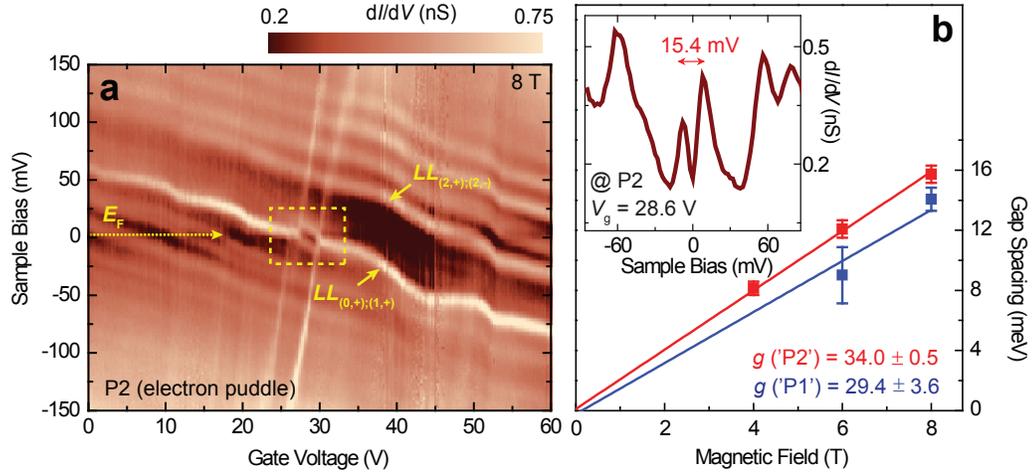

**Figure 5: Symmetry breaking in the $LL_{(0,+);(1,+)}$ quartet.** **a**, d$I$/d$V$ gate map at 8 T in electron puddle P2. $LL_{(0,+);(1,+)}$ is observed to split into two peaks opening a subgap when intersecting $E_F$ inside the dashed yellow rectangle. STS parameters: sample bias -200 mV, tunneling current 200 pA and root-mean-square modulation voltage 4 mV. **b**, The subgap energy vs. magnetic field for electron puddles P1 and P2. A linear fit of the gap vs. magnetic field yields the slopes of $(1.70 \pm 0.21)$ meV T$^{-1}$ and $(1.97 \pm 0.03)$ meV T$^{-1}$ for the positions P1 and P2, respectively. (Inset) d$I$/d$V$ spectra in the middle of the subgap of the electron puddle P2 at $V_g = 28.6$ V as marked in the gate map **a**, showing the subgap size of 15.4 meV at 8 T.

# Supporting Material for

## Microscopic Polarization of Bilayer Graphene

Gregory M. Rutter, Suyong Jung, Nikolai N. Klimov, David B. Newell, Nikolai B. Zhitenev*, and Joseph A. Stroscio*

* To whom correspondence should be addressed:
Email: nikolai.zhitenev@nist.gov (N. B. Z.), joseph.stroscio@nist.gov (J.A.S.)

**This pdf file includes:**

    **I.**    Numerical Data Analysis of Landau Levels
    References
    Figure S1

# Supplementary Material

## I. Numerical Data Analysis of Landau Levels

The low energy band structure of graphene consists of four bands; two bands meet at the charge neutrality point, and two higher energy bands related to the $A_2$-$B_1$ dimer bonds are offset by the interlayer energy $\gamma_1$[1]. We adopt a two-band model, which includes only the lowest energy bands, to quantitatively analyze the LLs spectra[2,3]. The analysis of LLs is limited to the low energy regime (less than 150 meV), which meets the requirement of the two-band model; $E < \gamma_1 = 0.377$ eV. We have found that this simple two-band model gives results within 5 % of those from the full four-band model[4]. Thus, the introduced error for using the simpler model is less than our experimental uncertainties. The simple analytic expression for the LL energies is derived from the two-band model as given by Eq. 1 in the main text, which we use to fit the LLs spectra with the potential energy asymmetry, $\Delta U$ and Fermi velocity, $v_F$ as free parameters.

Figure S1a shows the individual $dI/dV$ spectrum obtained for the P1 electron puddle at $V_g = 33$ V. Up to nine Landau levels are observed over the displayed spectra range in Fig. S1a. A fit of the LL energies with Eq. 1 yields an energy asymmetry, $\Delta U = (-34.8 \pm 2.4)$ meV [5], and Fermi velocity of $v_F = (1.00 \pm 0.01) \times 10^6$ m s$^{-1}$ as shown in Fig. 3d in the main text.

The band gaps qualitatively follow the separation between $LL_{(0,+);(1,+)}$ and $LL_{(2,+);(2,-)}$. We plot in Fig. S1b the difference in LL energies between $LL_{(2,+);(2,-)} - LL_{(0,+);(1,+)}$, which can be obtained directly from the gate maps (Fig. 3c). It is clear that the determined energy asymmetries from the fitting of the LLs with Eq. 1 (Fig. S1c, d) match the overall trend seen in the difference between LLs. Additionally, we have confirmed that using the Fermi velocity as a variable fit parameter only weakly affects the gap determination as demonstrated in Fig. S1c, d. Using the velocity as a fitting parameter, however, we observe a few percent variation in Fermi velocity as displayed in Fig. S1e, which can be ascribed to the velocity renormalization effects due to electron-electron interactions[6].

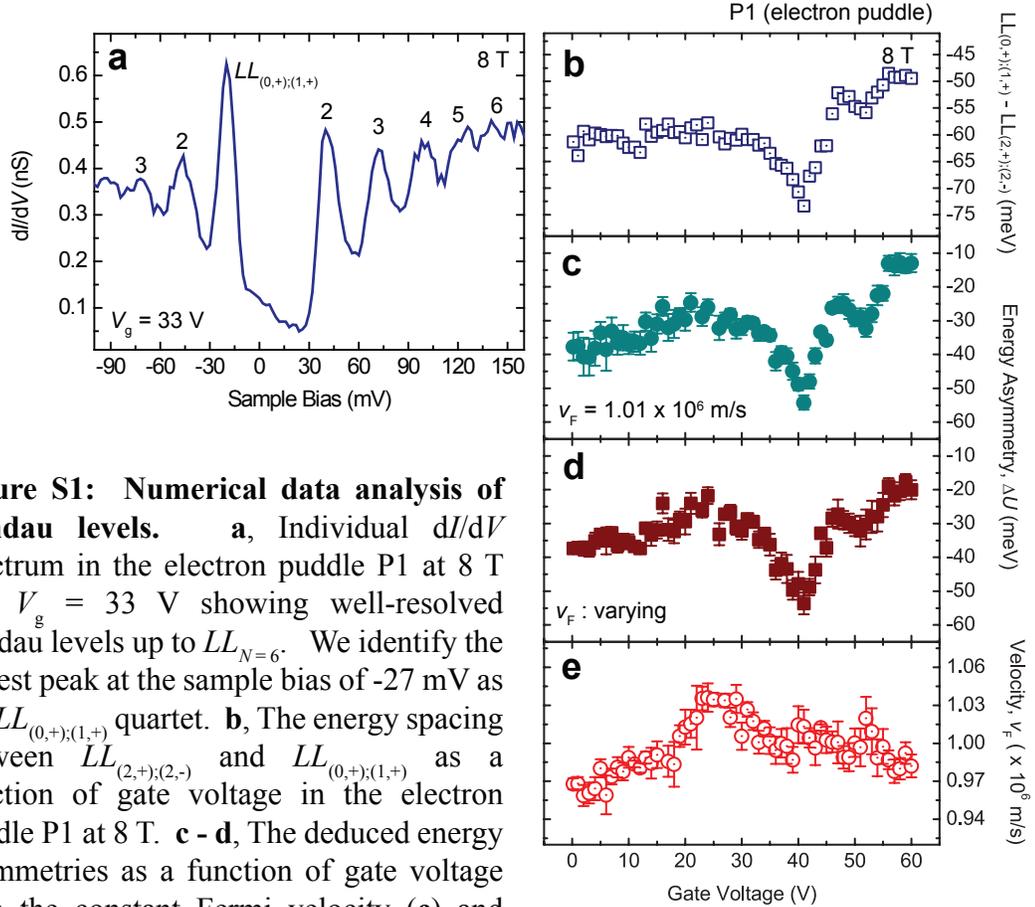

**Figure S1: Numerical data analysis of Landau levels.** **a**, Individual d$I$/d$V$ spectrum in the electron puddle P1 at 8 T and $V_g$ = 33 V showing well-resolved Landau levels up to $LL_{N=6}$. We identify the largest peak at the sample bias of -27 mV as the $LL_{(0,+);(1,+)}$ quartet. **b**, The energy spacing between $LL_{(2,+);(2,-)}$ and $LL_{(0,+);(1,+)}$ as a function of gate voltage in the electron puddle P1 at 8 T. **c - d**, The deduced energy asymmetries as a function of gate voltage with the constant Fermi velocity (**c**) and with the Fermi velocity as a fitting parameter (**d**). The band gaps follow the evolution of LL spacing displayed in **b**. **e**, The extracted Fermi velocity variation as a function of gate voltage from the LL fitting. The velocity renormalization, manifested by a few percent variations in magnitude does not affect the main features of bilayer energy asymmetry (Fig. S1**c**, **d**). Error bars one standard deviation, in (**c**) - (**e**).